\documentstyle[12pt]{article}
\def\singlespace {\smallskipamount=3.75pt plus1pt minus1pt
                  \medskipamount=7.5pt plus2pt minus2pt
                  \bigskipamount=15pt plus4pt minus4pt
                  \normalbaselineskip=15pt plus0pt minus0pt
                  \normallineskip=1pt
                  \normallineskiplimit=0pt
                  \jot=3.75pt
                  {\def\smallskip {\vskip\smallskipamount}}
                  {\def\medskip   {\vskip\medskipamount}}
                  {\def\bigskip   {\vskip\bigskipamount}}
                  {\setbox\strutbox=\hbox{\vrule
                    height10.5pt depth4.5pt width 0pt}}
                  \parskip 7.5pt
                  \normalbaselines}
\def\middlespace {\smallskipamount=5.825pt plus1.5pt minus1.5pt
                  \medskipamount=11.25pt plus3pt minus3pt
                  \bigskipamount=22.5pt plus6pt minus6pt
                  \normalbaselineskip=22.5pt plus0pt minus0pt
                  \normallineskip=1pt
                  \normallineskiplimit=0pt
                  \jot=5.825pt
                  {\def\smallskip {\vskip\smallskipamount}}
                  {\def\medskip   {\vskip\medskipamount}}
                  {\def\bigskip   {\vskip\bigskipamount}}
                  {\setbox\strutbox=\hbox{\vrule
                    height15.75pt depth6.75pt width 0pt}}
                  \parskip 7.25pt
                  \normalbaselines}
\def\dblspc {\smallskipamount=7.5pt plus2pt minus2pt
                  \medskipamount=15pt plus4pt minus4pt
                  \bigskipamount=30pt plus8pt minus8pt
                  \normalbaselineskip=30pt plus0pt minus0pt
                  \normallineskip=2pt
                  \normallineskiplimit=0pt
                  \jot=7.5pt
                  {\def\smallskip {\vskip\smallskipamount}}
                  {\def\medskip   {\vskip\medskipamount}}
                  {\def\bigskip   {\vskip\bigskipamount}}
                  {\setbox\strutbox=\hbox{\vrule
                    height21.0pt depth9.0pt width 0pt}}
                  \parskip 15.0pt
                  \normalbaselines}

\def\gm{\gamma }
\def\al{\alpha }

\def\eps{\epsilon }

\def\be{\begin{equation}}

\def\j-{\J_-}

\def\ee{\end{equation}}

\def\bearr{\begin{eqnarray}}
\def\bearrs{\begin{eqnarray*}}
\def\eearr{\end{eqnarray}}
\def\eearrs{\end{eqnarray*}}
\def\barr{\begin{array}}
\def\earr{\end{array}}
\def\p{\partial}

\def\o{\omega}

\def\non\non{\nonumber}
\def\nn8{\nonumber\\[15pt]}
\def\l{\left}
\def\r{\right}
\def\un{\underline}

\def\f{\frac}

\begin{document}
\input epsf
\middlespace
\begin{center}
{\Large{\bf Experimental tests of  curvature couplings of Fermions
 in General Relativity}}\\[30pt] S. Mohanty, B. Mukhopadhyay and A. R.
Prasanna\\
Physical Research Laboratory\\
Ahmedabad 380 009, India\\[40pt]

\un{Abstract}\\
\end{center}

Spin $\f{1}{2}$ particles in geodesic trajectories experience no
gravitational potential but they still have non-zero couplings to
the curvature tensor. The effect of space time curvature on
fermions can be parameterized by a vector and a pseudo-vector
potential. These apparent $CPT$ violating terms can be measured
with satellite based spin-polarized torsion balance and clock
comparison experiments. The Earth's curvature effect is of the
order of $10^{-37} Gev$ which is not far from the present bounds
of $\sim 10^{-29} Gev$ on such $CPT$ violating couplings.
\newpage

The effect of gravitational potential on quantum mechanics of
elementary particles was observed in the classic COW experiment
[1] using neutron interferometry. In general relativity it is
possible to choose a local inertial (free fall) frame where the
gravitational potential on an elementary particle is zero. The
curvature tensor in such frames however cannot be made to vanish
even locally. The experimental tests of this curvature coupling
would be an important test of general relativity. The couplings of
spin $1/2$ particle to curved space background have been studied
by many authors [2-9]. Parker [2] studied the effect of Riemann
curvature of the Schwarschild metric on the energy levels of
hydrogen atoms. It was found that the energy level shift on the
surface of the sun is of the order of $GM_\odot/(R^3 m_e)\sim
10^{-51} Gev$ which is too tiny to have any observational
consequence. The effect of rotation in the curved space metric on
spin $1/2$ particles have been studied in [3-9].

In this paper we show that the gravitational curvature couplings
can be parameterised as an external vector $\bar \psi \gamma^\mu
a_\mu \psi$ and a pseudo-vector $\bar \psi \gamma^5 \gamma^\mu
b_\mu \psi$ interaction term in the Lagrangian. In a rotating
inertial frame (like in the frame of a satellite orbiting the
earth) both $a_\mu$ and $b_\mu$ have some non-zero components while in a
non-rotating free fall frame only $a_\mu$ is non-zero.
From the phenomenological point of view a non-zero $b_\mu$ is of
interest as it can be measured in various experiments [11-18]
discussed below. The magnitude of $|\vec b|$ in earth orbiting
satellites is of the order of $6.5 \times 10^{-37} Gev$. This can
be compared with the sensitivity of experiments [12-14] like spin
polarised torsion balance and spin magnetisation measurement
 which can measure upto $|\vec b| \sim 10^{-29}
Gev$. The best prospects for measuring curvature effects on
fermions is by using macroscopic spin polariased substances in
earth orbit satellites.

 The general invariant coupling of spin $\f{1}{2}$ particles
to gravity is described by the Lagrangian [2-9]
\be
{\cal{L}} = \sqrt{-g} \l( \bar{\psi}i \gm^a D_a \psi - m \bar{\psi}
\psi \r) \ee where $$ D_a = e^\mu_a \l( \p_\mu - \f{i}{4}
\omega_{bc\mu} \sigma^{bc} \r), \eqno(2a) $$ $$
\sigma^{bc} = \f{i}{2} \l[ \gm^b, \; \gm^c \r] ,\eqno(2b) $$ $$
\omega_{bc\mu} = e_{b\lambda}\l(\p_\mu e^\lambda_{\;\;c} + \Gamma^\lambda_{\gamma
\mu} e^\gamma_{\;\;c}\r) $$
 where $a,b,c$ etc. denote flat space
indices and $\al$, $\beta$, $\gamma$ etc. are the curved space
indices. We use coordinates which are locally inertial along the
entire geodesic trajectory of the particles (called the Fermi
normal coordinates [10]). The metric in these coordinates to
second order takes the form $$ g_{00} = - 1 - R_{0\ell 0 m} X^\ell
X^m\eqno(3a) $$ $$ g_{0i} = g^{0i} = \f{2}{3} R_{0\ell im} X^\ell
X^m\eqno(3b) $$ $$ g_{ij} = \delta_{ij} - \f{1}{3} R_{i\ell jm}
X^\ell X^m\eqno (3c) $$ where $X^i$ are the spatial coordinates of
an event occurring at time $X^0$. The corresponding vierbeins are
given by $$ e^\al_{\;\; 0} = \delta^\al_{\;\; 0} - \f{1}{2}
R^\al_{\;\; \ell 0 m} X^\ell X^m\eqno(4a) $$ $$ e^\al_{\;\; i} =
\delta^\al_{\;\; i} - \f{1}{6} R^\al_{\;\; \ell im} X^\ell
X^m.\eqno (4b) $$ In this coordinate system Chritoffel connections
$\Gamma_{\mu\;\;\;\;\nu}^{\;\;\al} = 0$, on a geodesic, but their
first derivatives are non-zero and are related to the Riemann
tensor as given by $$ \Gamma_{\mu\;\;\;\; 0,\nu}^{\;\;\al} =
R_{\mu\;\;\;\; \nu 0}^{\;\; \al}\eqno(5a) $$ $$
\Gamma_{i\;\;\;\;j,k}^{\;\;\mu} = -\f{1}{3} \l(
R_{i\;\;\;\;jk}^{\;\;\mu} + R_{j\;\;\;\;ik}^{\;\;\mu} \r)\eqno(5b)
$$ \addtocounter{equation}{4} on any point along the geodesic.
Using the coordinates described in equations (3)-(5), it can be
shown, after some algebra, that the Dirac Lagrangian (1) can be
written in the form \be \barr{lll} \cal{L}& =& i \l( det \; e \r)
\bar{\psi} \l[\gm^0 \p_0+ \gm^i \p_i - a_0 \gm^0 - a_i \gm^i  -
b_0 \gm_5 \gm^0 - b_i \gm_5 \gm^i +i m \r] \psi \earr \ee (where
$i = 1,2,3$), the vector $\l( a_0 , \vec{a} \r)$ is given by $$
a_0 = \f{1}{4} R_{0\;\;im}^{\;\;i} X^m\eqno(7a) $$ $$ a_i =
\f{1}{2} R_{i00m} X^m - \f{1}{4} R_{i\;\;\;\;jm}^{\;\;j}
X^m\eqno(7b) $$ and the pseudovector $\l( b_0, \vec{b}_i\r)$ is
given by $$ b_0 = \f{1}{8} \eps_{0ijk} R^{jki}_{\;\;\;\;\;m}
X^m\eqno(8a) $$ $$ b_i = \f{1}{4} \eps_{0kji} R^{kj}_{\;\;\;\;0m}
X^m + \f{1}{4} \eps_{0kji} R^{0jk}_{\;\;\;\;\;m} X^m.\eqno(8b) $$
\addtocounter{equation}{2} From (6), we see that the effect of
space time curvature appears as an external vector coupling
$\gm^\mu a_\mu$ and pseudovector coupling $\gm^5 \gm^\mu b_\mu$
and is formally similar to the effective $CPT$ and Lorentz
violating Lagrangian [11-18].

Although the terms arising from gravitational curvature couplings
in (6) are the same as in the explicit $CPT$ violating
interactions studied in [11-18] there is a fundamental difference
between the two, in that, for the case of gravitational couplings
there is no $CPT$ of Lorentz symmetry violation. If one treats the
vectors $a_\mu$ and $b_\mu$ as fixed external vectors which do not
transform under $CPT$ (as is done in ref. [11-18]) then since both
$\bar \psi \gamma^\mu \psi$ and $\bar \psi \gamma_5 \gamma^\mu
\psi$ are odd under $CPT$ [19] the interaction terms $a_\mu \bar
\psi \gamma^\mu \psi$ and $b_\mu \bar \psi \gamma_5 \gamma^\mu
\psi$ explicitly violate $CPT$. Similarly the existence of
preferred external four vector $a_\mu$ and $b_\mu$ explicitly
violates the Lorentz invariance of vacuum. In the case of
gravitational couplings (6) however, it can be checked explicitly
that if one transforms the source currents which generate $a_\mu$
and $b_\mu$ correctly under $CPT$ then the interaction terms in (6) do not
violate $CPT$. One can check this explicitly from the expressions
for $a_\mu$ and $b_\mu$ given in (10) and (12) respectively for
the case of a satellite orbiting a central body with angular
velocity $\omega$. Under time reversal operation $T : \omega
\rightarrow -\omega$ and under parity $P$ the spatial three vector
$(X,Y,Z) \rightarrow (-X,-Y,-Z)$. One can see explicitly from the
expressions for $a_\mu$ (10) and $b_\mu$ (12) that under $CPT$
both $a_\mu$ and $b_\mu$ pick up  negative signs which compensate
for the negative signs picked up by the fermion bilinears  $\bar
\psi \gamma^\mu \psi$ and $\bar \psi \gamma_5 \gamma^\mu \psi$ under
$CPT$ transformation; and the gravitational curvature terms (6) do
not violate $CPT$ . Similarly since both $a_\mu$ and $b_\mu$
transform as four vectors under Lorentz transformation the
interaction terms in (6) are Lorentz invariant.  If one considers
the dynamics of fermions in a background gravitational field and
neglect the back-reaction of the fermions to the gravitating
sources, then the phenomenological effects of gravitational
curvature couplings in (6) will be the same as that of an external
fixed vector fields as studied in ref. [11-18] and the experiments
which can be used for testing $CPT$ violation using fermions as
test particles can also be used in principle for looking for
gravitational curvature couplings.

We consider a coordinate system in a free fall orbit around a
gravitating body. Such a system of coordinates would ideally
describe satellite based experiments. For Earth-based experiments,
the gravitational source would be the Sun. Choosing without loss
of generality orbital angular momentum around the Z-axis, $\vec\o
= \l( 0, 0, \o \r)$, we can write the non-zero components of the
Riemann tensor in the orbital free fall coordinates as follows:
\be
\barr{lll}
R_{1010}&=& \f{2GM}{R^3}\\[12pt]
R_{0202}&=& R_{0303} = - \f{GM}{R^3}\\[12pt]
R_{1212} &=& \f{GM}{R^3} \l( 1 + \o^2 \l( - Y^2 + 2X^2 \r) \r)\\[12pt]
R_{1313}&=& \f{GM}{R^3} \l( 1 - \o^2 Y^2 \r)\\[12pt]
R_{2323}&=& - \f{GM}{R^3} \l( 2 + \o^2 X^2 \r)\\[12pt]
R_{1202}&=& R_{1303} = - \f{GM}{R^3} \o Y\\[12pt]
R_{2101}&=& - \f{2GM}{R^3} \o X\\[12pt]
R_{2303}&=& \f{GM}{R^3} \o X\\[12pt]
R_{2313}&=& \f{GM}{R^3}\o^2 XY
\earr
\ee
where we have chosen $X$ along the radial direction and $Y$ along the
tangential direction of the orbit. Using the curvature components
(9), we can compute the vector $a_\mu$ and pseudovector $b_\mu$ given
in (7) and (8) for a free fall coordinate rotating with angular
velocity $\o$ around the $Z$-axis as follows. The vector couplings
$\bar{\psi} \gm^\mu a_\mu \psi$ of fermions is given by $a_\mu$
\be
\barr{lll} a_0&=& \f{3}{4} \f{GM\o}{R^3} XY\\[12pt] a_1&=& -\f{GM
X}{2R^3} + \f{GM\o^2 X}{4R^3} \l( 2X^2-Y^2 \r)\\[12pt]
a_2&=& \f{GMY}{4R^3} \l[ 1+\o^2 \l(2 X^2- Y^2\r) \r]\\[12pt]
a_3&=& \f{GMZ}{4R^3} \l[ 1 - \o^2 \l( X^2 + Y^2 \r) \r]. \earr \ee
The vector field $a_\mu$ can arise in
a Schwarzschild metric [2]. In a solar mass star it can split the
$2P_{3/2}$ levels of a hydrogen atom by amounts [2] \be \Delta E
\l( 2 P \r) \simeq \f{GM}{R^3}  \l( \f{1}{m_e} \r). \ee On the
surface of the Sun, this level split is 3.8$\times$10$^{-51} Gev$
and on the surface of Earth, the hydrogen $3P_{3/2}$ levels will
split by 1.2$\times$10$^{-51} Gev$. This energy is unobservably
small and at present there are no known experiments which can hope
to measure the gravity induced vector potential $a_\mu$.

The effective pseudo-vector couplings
$\bar{\psi} \gm^5 \gm^\mu b_\mu \psi$ are described by components
of $b_\mu$: \be (b_0,b_1,b_2,b_3)=(0,0,0,
 \f{GM\o}{R^3} X^2). \ee Much more stringent
bounds can be put on the pseudovector $b_\mu$ from CPT and Lorentz
violation tests [11-18]. The pseudo-vector term in the
non-relativistic limit is equivalent to the interaction energy
\be
H_I= - \vec s\cdot \vec b \ee due to the interaction of the
fermion spin $\vec s$ with the external field $\vec b$. In
experiments where a macroscopic number of fermions can be
polarised in the same direction, this interaction energy may be
measurable. In the Eot-Wash II experiment [12] , the
spin-polarised torsion balance has $N=8\times 10^{22}$ aligned
spins (with negligible net magnetic moment). There will be a
torque on such a torsion balance of the magnitude $\tau = (N/ \pi)
\Delta E$ where $\Delta E= |\vec b|$ is the energy difference
between the fermion spins polarised parallel and anti-parallel to
the external $\vec b $ field. From the results of this experiment
[12] it is possible to measure upto $|\vec b| \sim 10^{-28} Gev$
[13].

 Another method of probing $\vec b$ is to measure the net
magnetisation in a paramagnetic material using a squid [14]. An
external $\vec b$ field appears as an effective magnetic field of
strength $\vec B_{eff} = (\vec b /\mu_B)$. The magnitude of the
effective magnetic field which can be probed in this experiment is
$B_{eff} =10^{-12} gauss$ which translates to a measurent of the
$\vec b$ field at the level of $10^{-29} Gev$ [13,14].

 Bounds on spatial
components of $b_\mu$ can also be put from muon properties [15],
tests of QED in Penning traps [16]
 spectroscopy of hydrogen and
anti-hydrogen  [17] and in future clock-comparison tests with
satellite based atom clocks [18].

We have shown above that a non-zero $b_\mu$ arises due to
curvature couplings of fermion in a rotating frame.  On a satellite
orbiting the Earth with a typical velocity of 7.5 km/sec the value
of $b_3 = \l( \f{GM_\oplus}{R_\oplus} \r)\; \o =$
6.5$\times$10$^{-37} Gev$ . This is a factor of 10$^{-6}$ smaller
than the best available bounds at present, but it may be possible
to measure curvature effects in future satellite based
experiments. If one considers the Earth's motion around the Sun,
$b_3= \l( \f{GM_\oplus}{R_{ES}} \r) \o = 1.2 \times 10^{-40} GeV$
is much smaller. On the Mercury orbit around the Sun, taking
Mercury-Sun distance $\sim 55 \times 10^6 km$  and period of
Mercury orbit $\sim 0.24$ years, $b_3 = 2.1\times10 ^{-39} Gev$.
These estimates show that the best prospect of observing the
curvature effects on fermions is probably by spin-polarised
torsion balance experiments  in  low orbit Earth satellites.

There can be some interesting cosmological application of
gravitational curvature couplings in early universe where
curvature effects of the expanding universe is likely to be large.
It has been shown by Bertolami et al. [20] that explicit $CPT$
violating terms in the Lagrangian can give rise to a net chemical
potential for baryon number and generate a net baryon asymmetry in
the presence of baryon number violating processes. In a separate
paper [21] we show that $T$ violation induced by the expanding
universe R-W metric gives rise to a $CP$ violating interaction
term in the Lagrangian which can give rise to net baryon asymmetry
in the present universe.

\newpage
\begin{center}
{\bf References}
\end{center}
\begin{enumerate}

\item R. Colella, A. W. Overhauser and S. A. Werner (1975) {\it Phys. Rev. Lett.} {\bf
34}, 1472.
\item L. Parker (1980) {\it Phys. Rev. Lett.} {\bf 44}, 1559;\\
 L. Parker (1980) {\it Phy. Rev.} {\bf D22}, 1922.
\item F. W. Hehl and W. T. Ni (1980) {\it Phy. Rev.} {\bf 42}, 2045.
\item E. Fischbach, B. S. Freeman and W. K. Cheng (1981) {\it Phys. Rev.} {\bf D23 },
2157.
\item D. Choudhary, N. D. Hari Dass and M. V. N. Murthy (1989) {\it
Class. Quan. Grav.} {\bf 6}, L167.
\item C. Q. Xia and Y. L. Wu (1989) {\it Phys. Lett.} {\bf A141}, 251.
\item Z. Lalak, S. Pokorski and J. Wess (1995), {\it Phys. Lett.} {\bf B
355}, 453.
\item S. Wajima, M. Kasai and T. Futamase (1999) {\it Phys. Rev.} {\bf D 55},
1964.
\item  K. Varju and L. H. Ryder (2000) {\it Phys. Rev.} {\bf D 62}, 024016.
\item F. K. Manasse and C. W. Misner (1963) {\it J. Math. Phys.} {\bf 4}, 735.
\item D. Colladay and V. A. Kostelecky (1997), Phys. Rev. {\bf D55},
6760;\\
V. A. Kostelecky (Ed.) (1999) {\it CPT and Lorentz Symmetry}, World
Scientific, Singapore.
\item E. G. Adelberger et al. in P. Herczag et al. Eds. (1999) ,
{\it
Physics beyond the standard model.} p717, World Scientific,
Singapore.
\item R. Bluhm and V. A. Kostelecky (2000) {\it Phys. Rev. Lett.} {\bf
84}, 1381.
\item W.-T. Ni et al. (1999) {\it Phys. Rev. Lett.} {\bf 82}, 2439.
\item R. Bluhm, V. A. Kostelecky and C. D. Lane (2000) {\it Phys. Rev. Lett.}
{\bf 84}, 1098.
\item H. Dehmelt et al. (1999) {\it Phys. Rev. Lett.} {\bf 83}, 4694;\\
G. Gabrielse et al. (1999) {\it Phys. Rev. Lett.} {\bf 82}, 3198;\\ R. Bluhm,
V. A. Kostelecky and N. Russel (1997) {\it Phys. Rev. Lett.} {\bf
79}, 1432.
\item R. Bluhm, V. A. Kostelecky and N. Russel (1999) {\it Phys. Rev. Lett.}
{\bf 82}, 2254.
\item R. Bluhm et al. (2000) {\it Clock-Comparison Tests of Lorentz
and CPT Symmetry in Space}, hep-ph/0111141.
 \item C. Itzykson and J-B Zuber (1985) {\it Quantum Field Theory} , Pg
 157, McGraw Hill Book Co.- Singapore.
 \item O. Bertolami, D. Colladay, V. A. Kostelecky and R. Potting
 (1997), Phys. Lett. {\bf B 395}, 178.
\item S. Mohanty, B. Mukhopadhyay and A. R. Prasanna, {\it T
violation by curvature couplings of fermions in early universe and
baryogenesis.} under preparation.
\end{enumerate}
\end{document}